# Atomic excitation during recollision-free ultrafast multi-electron tunnel ionization

W. A. Bryan[1,3,6,*], S. L. Stebbings[1,4], J. McKenna[2], E. M. L. English[1], M. Suresh[2,5], J. Wood[1], B Srigengan[2], I. C. E. Turcu[3], J. M. Smith[3], E. J. Divall[3], C. J. Hooker[3], A. J. Langley[3], J. L. Collier[3], I. D. Williams[2] and W. R. Newell[1]

[1]*Department of Physics and Astronomy, University College London, Gower Street, London WC1E 6BT, UK*

[2]*Department of Pure and Applied Physics, Queen's University Belfast, Belfast BT7 1NN, UK*

[3]*Central Laser Facility, CCLRC Rutherford Appleton Laboratory, Chilton, Didcot, Oxon. OX11 0QX, UK*

[4]*Present address: Department of Physics and Astronomy, University of Southampton, Southampton, SO17 1BJ, UK*

[5]*Present address: Cavendish Laboratory, University of Cambridge, Madingley Road, Cambridge CB3 0HE, UK*

[6]*Now at: Physics Department, Swansea University, Singleton Park, Swansea SA2 8PP, UK*

**To whom correspondence should be addressed: w.a.bryan@swansea.ac.uk*

**Modern intense ultrafast pulsed lasers generate an electric field of sufficient strength to permit tunnel ionization of the valence electrons in atoms[1]. This process is usually treated as a rapid succession of isolated events, in which the states of the remaining electrons are neglected[2]. Such electronic interactions are predicted to be weak, the exception being recollision excitation and ionization caused by linearly-**

**polarized radiation[3]. In contrast, it has recently been suggested that intense field ionization may be accompanied by a two-stage 'shake-up' reaction[4]. Here we report a unique combination of experimental techniques[5 - 8] that enables us to accurately measure the tunnel ionization probability for argon exposed to 50 femtosecond laser pulses. Most significantly for the current study, this measurement is independent of the optical focal geometry[7,8], equivalent to a homogenous electric field. Furthermore, circularly-polarized radiation negates recollision. The present measurements indicate that tunnel ionization results in simultaneous excitation of one or more remaining electrons through shake-up[9]. From an atomic physics standpoint, it may be possible to induce ionization from specific states, and will influence the development of coherent attosecond XUV radiation sources[10]. Such pulses have vital scientific and economic potential in areas such as high-resolution imaging of in-vivo cells and nanoscale XUV lithography.**

The Noble gas atoms are a natural choice for studying electron dynamics in an ultrafast intense laser pulse, and a number of pivotal experimental studies have established this rich and vibrant field of research, revealing much about the complex nature of photoionization[11]. However, the majority of these studies have employed linearly polarized radiation, and used the predictions of tunnelling theory[2] (generally the ADK treatment[12]) in which the target is confined to the ground state. The current investigation was stimulated by the theoretical work of Eichmann *et al*[13] and Zon *et al*[4], and reveals experimental evidence for a hereto unobserved accompanying excitation mechanism.

In the ultrafast regime, the optical pulse duration is of the order of femtoseconds (1 femtosecond = $10^{-15}$ s) and ionization proceeds either by a multiphoton perturbative process[14] or a nonperturbative strong-field process, generally accepted as being described by tunnel theory: for a recent review see Popov[15]. The present work is concerned purely with strong field processes (peak intensity greater than approximately 10 TW/cm$^2$, where 1 TW = $10^{12}$ W) where the rate of tunnel ionization is governed by the frequency and strength of the radiation-induced electric field and the binding energy and quantum state of the ion and electron(s).

Immediately following ionization, the 'drive pulse' electron is in a Volkov state[16] and initially is fully correlated with the parent ion. As the electron is accelerated by the laser field, the electron trajectory is determined by the ellipticity of the field and the phase at which ionization occurred. Electron impact excitation[17] or subsequent ionization[18] can arise in a linearly polarized laser field through the intriguing phenomenon of recollision. Recollision is the key mechanism for attosecond XUV pulse generation, as the kinetic energy of the electron is dissipated photonically if the electron is recaptured by the parent ion[19]. However, in the present work, we make recollision events negligible by employing circularly polarized light: the absorption of a large number of photons transfers considerable angular momentum to the liberated electron, preventing it from returning to the ionic core[7, 20], thus the masking effect of recollisional excitation and ionization are negated.

While strong-field ionization and recollision in ultrafast laser pulses is well documented, minimal theoretical and essentially no experimental studies have investigated the possibility of simultaneous excitation of the parent ion during tunnel ionization. In the case of multiphoton ionization, resonant excitation processes have been identified[21 - 23], however in the present work, the intensity of the laser field is sufficient that tunnel ionization occurs with a dramatically enhanced probability. The contemporary work of Zon[4,24,25] invoked the idea of 'inelastic tunnelling' whereby the parent ion is left in an excited state following the ionization of one of *N* identical valence electrons. Excitation is through 'shake-up', first employed by Carlson[9] to explain single UV photon absorption leading to the ionization of a first and the excitation of a second electron. The ionization event diabatically distorts the electron wavefunctions, resulting in the excitation of a bound second electron. Using this concept, Zon *et al*[4,24] and Kornev *et al*[25] have derived a general expression for the rate of tunnel ionization (TI) of an atom with simultaneous excitation of the lowest lying ionic states. This treatment relies on the sudden approximation (SA), which is valid provided accessing the excited states in $Ar^{(q+1)+}$ requires considerably less energy than the ionization potential of $Ar^{q+}$. Kornev *et al*[25] show that while this has negligible influence on the ionization probability for the $Ar^{q+} \rightarrow Ar^{(q+1)+}$ process, the creation of excited states in the product $Ar^{(q+1)+}$ ion strongly influence the $Ar^{(q+1)+} \rightarrow Ar^{(q+2)+}$ ionization probability. The consequence of this mechanism is illustrated in figure 1. Furthermore, by allowing for all combinations of excitation and ionization, the significant role of simultaneous excitation and ionization during tunnelling in the laser field is highlighted. To distinguish from standard sequential tunnel ionization (TI), we refer to such processes as multi-electron tunnel ionization (METI).

The analogous process of 'shake-off' resulting in further ionization rather than discrete excitation was originally proposed a decade ago by Fittinghoff *et al*[26] in an attempt to quantify the many-orders of magnitude discrepancy between the predicted sequential TI yield and the observed $He^{2+}$ ion yield by laser radiation[27]. It has been through careful experiments employing linear and circular polarization (see for example the work of Guo *et al*[28]) and the COLTRIMS technique[17,18] that recollision has finally been accepted as the mechanism responsible. More recently, the shake-off ionization of helium exposed to ultrafast laser pulses has been theoretically investigated using the *S*-matrix technique. Becker and Faisal calculated the ratio of the rates of shake-off to recollision, and demonstrated that shake-off was many orders of magnitude weaker[29]. As a result, the presence of shake-off in the strong-field ionization of atoms has been generally dismissed for the case of linear polarization. In the present work, we propose that the relevance of shake-up excitation be re-examined in the case of argon, given the validity of the SA is significantly different from the case of helium discussed above. The far lower-lying excited states in $Ar^{q+}$ (q = 1 to 6) are far more readily accessed during TI, in contrast to the extremely high excitation energies of $He^{+}$, irreconcilable with the SA.

In general, previous experimental measurements of ion yield as a function of laser intensity are a convolution of the ion signal with the focal volume producing the signal. By simply changing the energy of the laser pulse, the spatial distribution of laser intensity also changes. Frequently, the complexity of this situation is compounded by diffraction associated with the spatial profile of the laser[7]; a direct comparison with theory is thus impossible without introducing the specific experimental geometry. A novel solution to this 'volume variation' problem, referred to as intensity selective scanning (ISS), was demonstrated by Van Woerkum[5]; the relevance of ISS is described in *Methods*, and is illustrated in figure 2. The ion yield recorded at each position $z_f$, the distance of the spectrometer axis from the focus, still inherently depends on the intensity distribution in the $(x, y)$ plane. Two numerical methods have been proposed for removing spatial integration over each 'slice' through the focal volume: Walker *et al*[5] described a deconvolution routine, and Goodworth *et al*[8] proposed a tomographic method. Both recover focal geometry-independent probabilities of ionization; the authors have extended the former to experimentally realistic non-Gaussian focal conditions[7].

In figure 2, we present a measurement of the spatial distribution of ions generated when argon is exposed to circularly polarized 50 fs pulses over the intensity range 100 TW/cm$^2$ to 100 PW/cm$^2$. This experiment was performed using the ASTRA Laser Facility at the Rutherford Appleton Laboratory, as detailed in *Methods*. The $Ar^{q+}$ ($q$ = 1 to 6) ion yield was measured as the laser focus was translated by a computer-driven motion stage, exposing the detector to different 'slices' through the focal volume: the spatial selectivity of the time-of-flight mass spectrometer (TOFMS) (grey arrow) is indicated by the narrow slice in the (*x,y*) plane.

The partial probability of ionization PPI to $Ar^{q+}$ ($q$ = 1 to 6) shown in figure 3 is recovered from the ISS data (figure 2) using the technique pioneered by the authors[7]. Importantly, we totally remove any instrument dependence, including the enhancement of detector gain with charge state. The distinction 'partial' is necessary as the deconvolution method is only valid at intensities below saturation (as indicated by the dashed line), irrespective of diffraction effects. The saturation intensity is defined as described in *Methods*. The PPI for each ionization state (figure 3) can be directly recast in terms of the universal conserved probability of ionization (CPI), which embraces all charge states. The CPI for each charge state is presented in figure 4, and is discussed in *Methods*.

The spatially homogenous intensity-dependent predictions of Kornev *et al*[25] are perfect for direct comparison with the measured CPI data. Such a comparison is presented in figure 4(a)-(f), where the data points are the present measurements, the thin lines the calculated probabilities of producing $Ar^{q+}$ from the $Ar^{(q-1)+}$ ground state only (sequential TI), and the thick lines are the METI predictions including shake-up. Note the experimental CPI has been normalized to the theoretical predictions via the rising edge of the $Ar^{+}$ CPI; this first ionization stage will not be influenced by excitation. Figure 4 clearly highlights the significant difference between theory and experiment when only sequential TI from the ground state is considered (thin line), lying generally well outside the experimental uncertainty. This is illustrated in figure 4 by representative error bars. However, when shake-up is included in the METI prediction, an excellent agreement is found for all charge states as indicated by the thick lines in figure 4. While other possible excitation mechanisms could also be considered, for example the ‘way-out’ excitation, whereby the departing electron scatters from another bound electron[3], the observed agreement is strongly supportive of the shake-up model. This benefits from the repeated diabatic distortion of the valence electronic wavefunctions during tunnel ionization at high intensities over a number of optical cycles at intensities greater than 1 PW/cm$^2$. Thus is an important contrast to single photon shake-off[30], as the cumulative influence of the oscillatory laser field need be considered.

To conclude, we have recorded for the first time strong evidence for the presence of considerable atomic excitation during tunnel ionization by a 790 nm 50 fs circularly polarized laser pulse focused to intensities in excess of 100 TW/cm$^2$. Such an observation is made without the need to embrace recollision processes. The significance of the agreement between our experimental observations and recent theoretical predictions indicate that excitation during ionization must be considered irrespective of recollision processes, referred to as multi-electron tunnel ionization. Such excitation will also occur in a linearly polarized laser field, and is expected to be even more influential in a 5 fs few-cycle laser pulse[25]. As the proposed method for generating intense bursts of XUV radiation relies on accurately controlling high harmonic generation in few-cycle intense laser fields[19], excitation during tunnelling has a major bearing on the emerging field of optical attosecond physics. Accurate manipulation of the liberated electron motion during the infrared ‘drive pulse’ and the occupation of initial and *transient* electronic states will allow tuning of the spectrum of the XUV pulse.

## Methods

**Intensity Selective Scanning.** We are interested in ionization and excitation processes that saturate at intensities less than 100 PW/cm$^2$, therefore the 30 mJ 790 nm 50 fs laser pulses generated by the ASTRA Laser Facility (UK) need only be softly focussed (f/11 optics) to generate a peak intensity in excess of 100 PW/cm$^2$. Indeed, multiple ionization occurs over tens of millimetres. By measuring the $Ar^{q+}$ ($q$ = 1 to 6) ion yield with a tightly apertured (250 microns) ion time-of-flight mass spectrometer, only those ions generated within a narrow spatial (and therefore intensity) window are detected. Then, as the focusing optic is translated, the spectrometer is only exposed to ions generated by a well-defined laser intensity[5-8, 25], as illustrated in figure 2. Throughout this measurement, the argon gas pressure is low enough so as to avoid space-charge effects, tested by repeating the ISS measurements as a function of target gas pressure. The narrowing of the spatial distribution of ions as a function of $z_f$ for increasing n is representative of the increasing ionization potential, where the highest ionization state presented ($Ar^{6+}$) is peaked at $z_f = 0$ mm where the intensity is at a maximum.

**Deconvolution of a non-Gaussian focus.** The ion yield measured by intensity selective scanning depends very specifically on the spatial distribution of intensity within our focal volume. To make this measurement more universal, we remove this dependence through a deconvolution technique[7], requiring the measured ion yield and theoretical on-axis intensity as a function of focal position as inputs. The resulting partial probability of ionization (PPI) is independent of the variation of the signal-producing volume. We are also able to account for the unavoidable diffraction of the laser pulse[7] by solving the Huygens-Fresnel diffraction integral. The PPI is equivalent to the response of a single atom to a spatially infinite laser focus, directly comparable to theoretical predictions, which tend to be presented in terms of a homogenous laser field. The PPI results are valid up to saturation, however at higher intensities the deconvolution breaks down, and the PPI is defined as unity. By measuring the first derivative of the PPI, the saturation intensity is defined as the point at which the gradient falls below the equivalent uncertainty on the PPI. The uncertainty is estimated by calculating the statistical deviation in the recorded average time-of-flight signal, and the upper and lower confidence interval corresponding to 1 s.d. was propagated through the deconvolution procedure. Furthermore, the probability at which saturation occurs depends on the quantum efficiency of the detector, which varies with charge state. The deconvolution method can be used to remove this dependence by normalising the saturation PPI to unity.

**Conservation of probability.** Consider the PPI($n$) for $n$ = 1, 2 as shown in figure 3. At low intensity, PPI(1) is small, and increases with intensity up to saturation at an intensity of 1 PW/cm$^2$. However, as is often the case in atomic ionization, PPI(2) is nonzero at this intensity. The condition for conserving probability is that the sum of probabilities is less than unity below the saturation intensity of PPI(1) or equal to unity above the saturation intensity of PPI(1), thus at an intensity greater than ~1 PW/cm$^2$, the conserved probability of ionization to $Ar^+$, CPI(1) must be less than unity as PPI(2) is nonzero. This definition is extended to an $N$ electron system in the present work. The uncertainty in the CPI is estimated by propagating the uncertainty in the PPI through the deconvolution procedure.

**Acknowledgements.** This work is funded by the Engineering and Physical Sciences Research Council (EPSRC), UK. Research studentships are acknowledged by J.W, E.M.L.E and S.L.S (EPSRC), J.McK (DEL) and M.S. (IRCEP at QUB). The authors gratefully acknowledge A. S. Kornev and B. A. Zon from Voronezh State University, Russia for fruitful discussions, suggestions and electronically communicating their theoretical data.

**Author Information.** Reprints and permissions information is available at npg.nature.com/reprintsandpermissions. The authors declare that they have no competing financial interests. Correspondence and requests for materials should be addressed to W.A.B. (w.bryan@ucl.ac.uk), I.D.W. (i.williams@qub.ac.uk) or W.R.N. (w.r.newell@ucl.ac.uk).

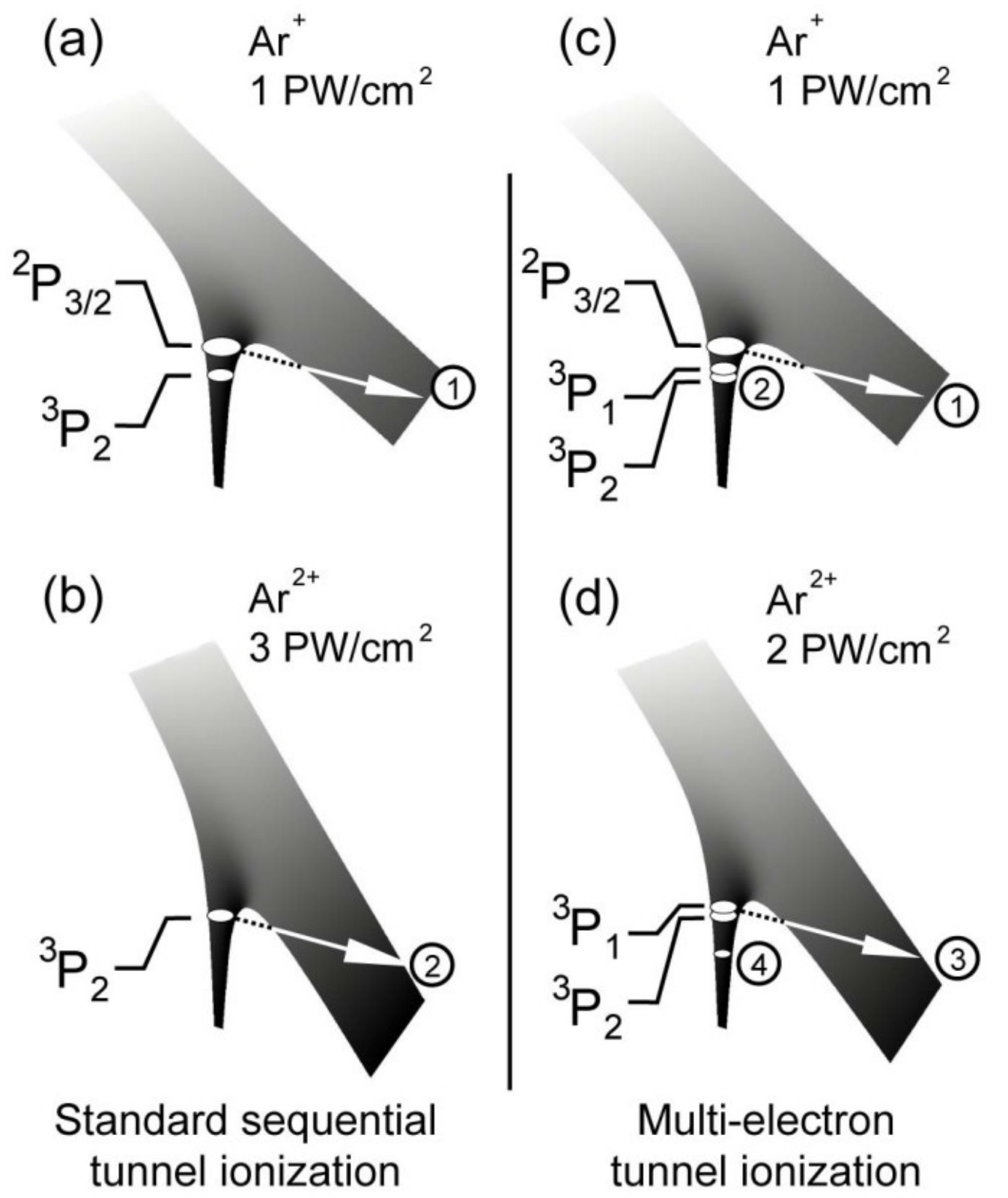

**Figure 1** Tunnel ionization and resulting excitation mechanisms in the Coulomb potential of argon. Regular sequential tunnel ionization: **a.** at an intensity of 1 PW/cm$^2$ the ground state ($^2P_{3/2}$) of $Ar^+$ has a significant probability to tunnel ionize (1). **b.** As the intensity is increased to 3 PW/cm$^2$, the ground state ($^3P_2$) of $Ar^{2+}$ can ionize (2). Considering multi-electron processes, the tunnel ionization is significantly influenced by excitation. **c.** As the $^2P_{3/2}$ electron tunnels from the $Ar^+$ (1), a significant population is transferred from the $^3P_2$ to $^3P_1$ excited state in the $Ar^{2+}$ ion (2). **d.** A lower intensity (2 PW/cm$^2$) is then required to ionize population in the $^3P_1$ state (3), while the $^3P_2$ state ionizes at 3 PW/cm$^2$ (**b**). The generation of $Ar^{2+}$ can then further excite the $Ar^{3+}$ ion (4). We measure the dependence of the probability of ionization on the laser intensity and charge state.

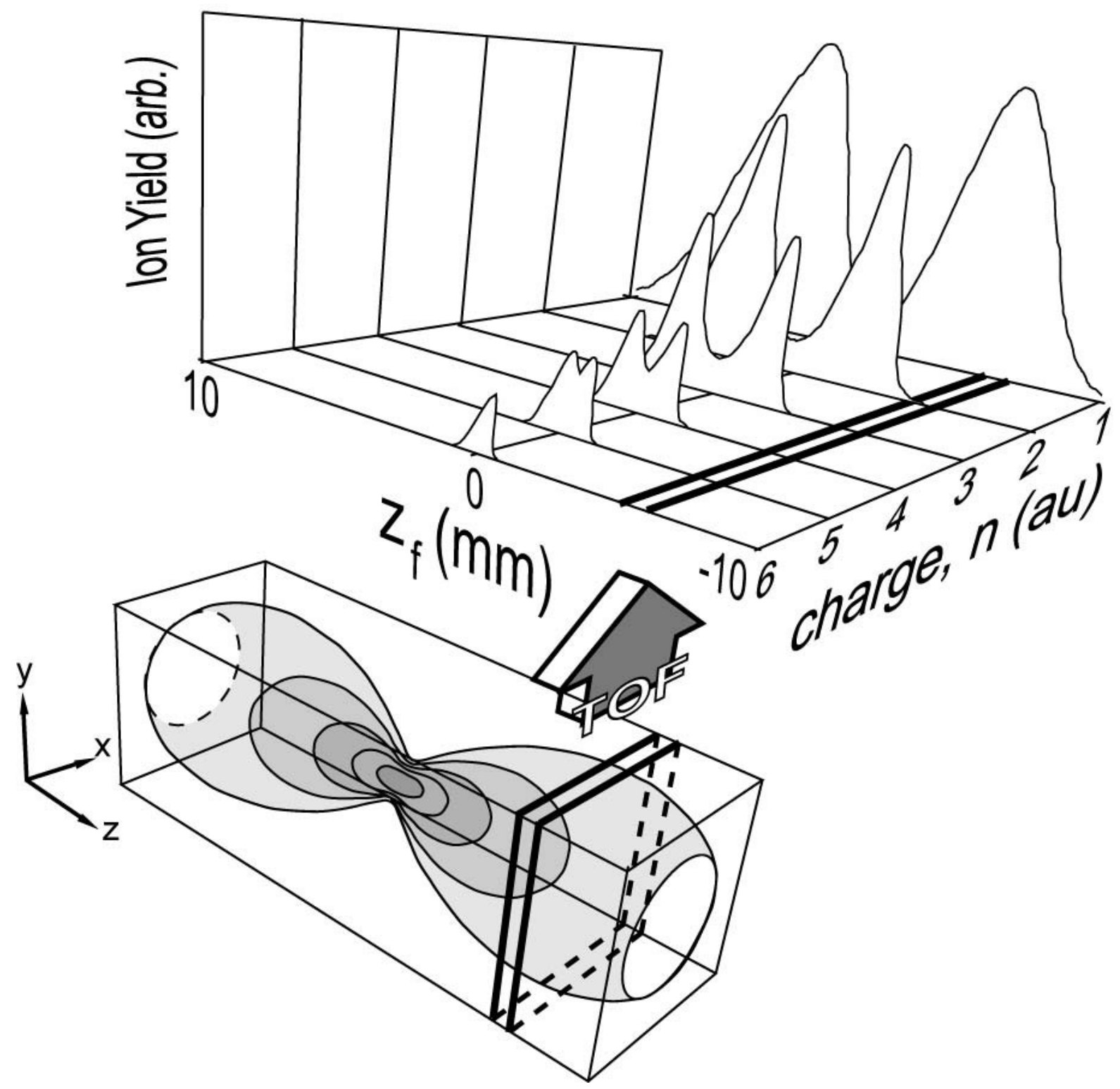

**Figure 2** Illustration of the intensity selective scanning (ISS) technique[5 - 8] and measured ion yield data. As the laser focus (bottom) is translated with respect to the narrow detection aperture of the TOFMS (slot perpendicular to $z_f$, the distance from the focus), the ion yield for all charge states of $Ar^{q+}$ (*q = 1 to 6*) is measured. The laser pulse propagates in the z direction, and the focus is translated parallel to this axis.

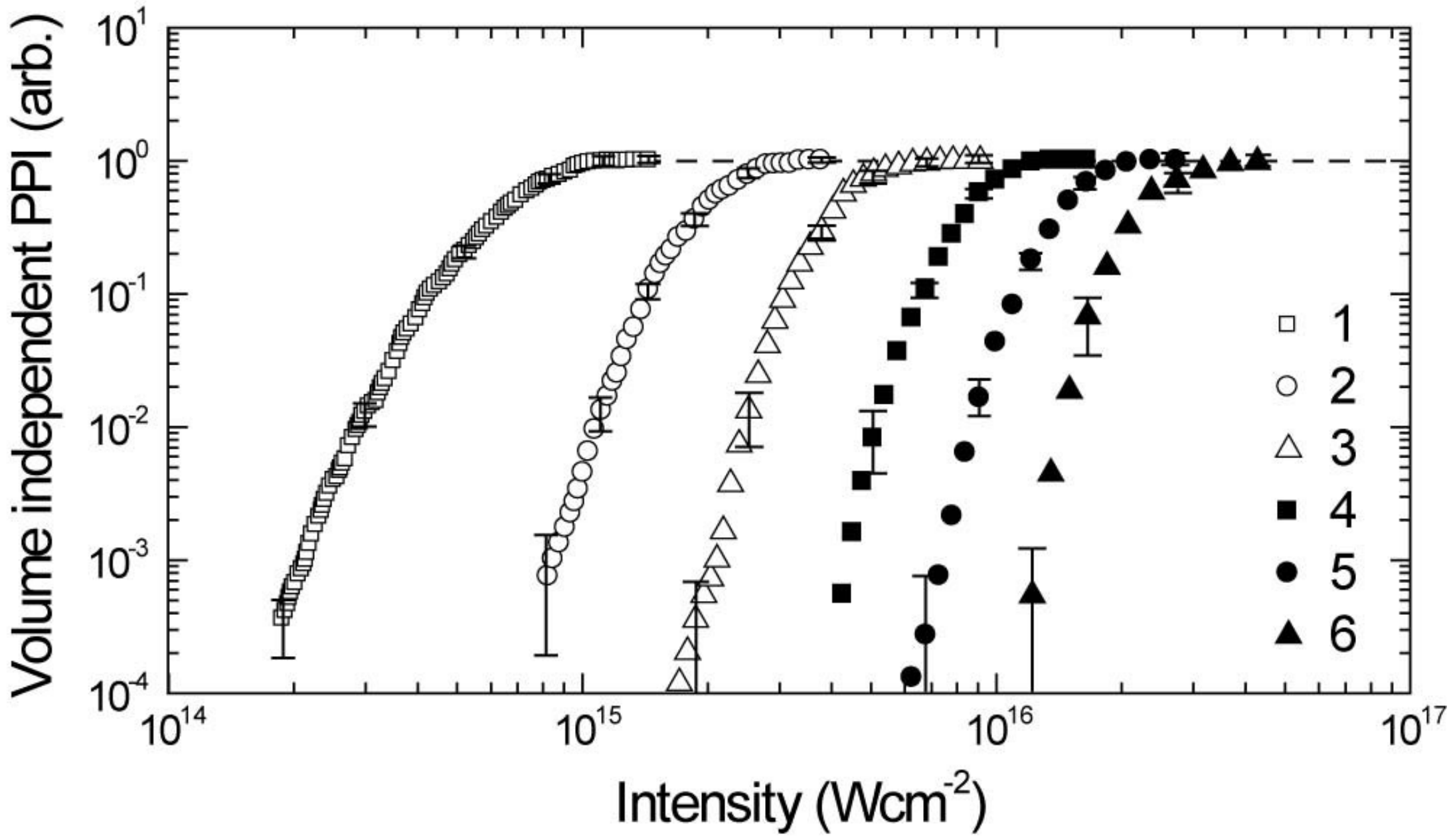

**Figure 3** Partial probability of ionization (PPI) to $Ar^{q+}$ ($q$ = *1 to 6*) as a function of spatially independent laser intensity. As circularly polarized radiation is employed, this data is free from the influence of recollision effects. The partial probability is recovered from the ISS data in figure 2 using the method described in reference [7]. The uncertainty is estimated by calculating the statistical deviation in the recorded average time-of-flight signal, and the upper and lower confidence interval corresponding to 1 s.d. was propagated through the deconvolution procedure. The dashed line indicates a PPI of unity, whereby ionization is saturated.

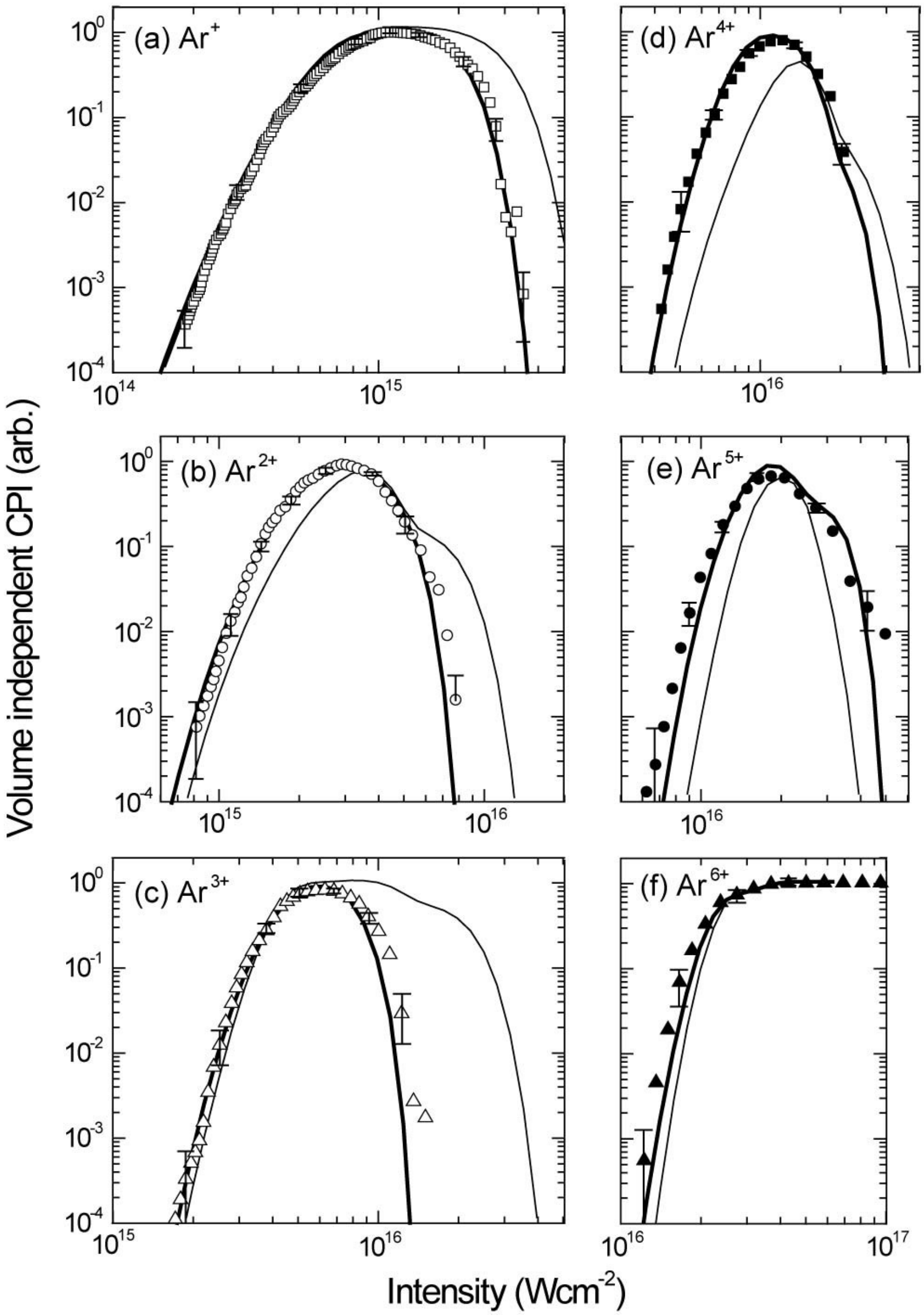

**Figure 4** Conserved probability of ionization (CPI) to $Ar^{q+}$ (*q = 1 to 6*) as a function of spatially-independent laser intensity. (a) $Ar^{+}$, (b) $Ar^{2+}$, (c) $Ar^{3+}$, (d) $Ar^{4+}$, (e) $Ar^{5+}$ and (f) $Ar^{6+}$. Data points: current results derived[7] from the data in figure 3. Thin lines: theoretical predictions[25] of sequential multiple tunnel ionization (TI) from ionic ground states only, equivalent to ADK theory[12], thick lines: full multi-electron tunnel ionization (METI) treatment allowing for shake-up

excitation during tunnelling. The uncertainty in the CPI is estimated by propagating the uncertainty in the PPI through the deconvolution procedure.